\documentclass[prl,twocolumn,superscriptaddress,showpacs]{revtex4}

\usepackage{graphicx}

\begin{document}

\newcommand{\ket}[1]{\mbox{$|\!#1\;\!\rangle$}}
\newcommand{\aver}[1]{\mbox{$<\!#1\!\!>$}}
\def\ua{\uparrow}
\def\da{\downarrow}

\title{Zeeman energy and spin relaxation in a one-electron quantum dot}

\author{R. Hanson, B. Witkamp, L.M.K. Vandersypen, L.H. Willems van Beveren, J.M. Elzerman and L.P. Kouwenhoven}
\affiliation{Department of NanoScience and ERATO Mesoscopic
Correlation Project, Delft University of Technology, PO Box 5046,
2600 GA Delft, The Netherlands}

\date{\today}

\begin{abstract}
We have measured the relaxation time, $T_{1}$, of the spin of a single electron confined in a semiconductor quantum dot (a proposed quantum bit). In a magnetic field, applied parallel to the two-dimensional electron gas in which the quantum dot is defined, Zeeman splitting of the orbital states is directly observed by measurements of electron transport through the dot. By applying short voltage pulses, we can populate the excited spin state with one electron and monitor relaxation of the spin. We find a lower bound on $T_{1}$ of 50 $\mu$s at 7.5 T, only limited by our signal-to-noise ratio. A continuous measurement of the charge on the dot has no observable effect on the spin relaxation.
\end{abstract}
\maketitle

The spin of an electron confined in a semiconductor quantum dot (QD) is a promising candidate for a scalable quantum bit \cite{loss,Lieven}. The electron spin states in QDs are expected to be very stable, because the zero-dimensionality of the electron states in QDs leads to a significant suppression of the most effective 2D spin-flip mechanisms \cite{Khaetskii1}. Recent electrical transport measurements of relaxation between spin triplet and singlet states of two electrons, confined in a pillar etched from a GaAs double-barrier heterostructure (``vertical'' QD), support this prediction (relaxation time $>\!$ 200 $\!\mu$s at $T\leq\!$ 0.5 K) \cite{FujisawaNature}. However, the triplet-to-singlet transition, in which the total spin quantum number \textit{S} is changed from 1 to 0, is forbidden by a selection rule ($\Delta S$=0) that does not hinder relaxation between Zeeman sublevels (which conserves \textit{S}). Therefore, measurements on a single electron spin are needed in order to determine the relaxation time of the proposed qubit. 

Relaxation between Zeeman sublevels in closed GaAs QDs is expected to be dominated by hyperfine interaction with the nuclei at magnetic fields below 0.5 T \cite{Siggi} and by spin-orbit interaction at higher fields \cite{Khaetskii2}. At 1 T, theory predicts a $T_1$ of 1 ms in GaAs \cite{Khaetskii2}; at fields above a few Tesla, needed to resolve the Zeeman splitting in transport measurements, no quantitative estimates for $T_1$ exist.

For comparison, in $n$-doped self-assembled InAs QDs containing one resident electron, pump-probe photoluminescence measurements gave a single-electron spin relaxation time of 15 ns (at $B$=0 T, $T$= 10 K) \cite{Cortez}. In undoped self-assembled InAs QDs, the exciton polarization is frozen throughout the exciton lifetime, giving a relaxation time $>$20 ns \cite{Paillard}. 

Electrical measurements of the single-electron spin relaxation time have up to now remained elusive. In vertical QDs, where electrical measurements on a single electron were reported almost a decade ago \cite{LeoFewEl}, it has been difficult to directly resolve the Zeeman splitting of orbitals \cite{ZeemanVQD}. Recently, the one-electron regime was also reached in single \cite{Ciorga} and double lateral GaAs QDs \cite{Jero}, which are formed electrostatically within a two-dimensional electron gas (2DEG) by means of surface gates. 

In this Letter we study the spin states of a one-electron lateral QD directly, by performing energy spectroscopy and relaxation measurements. We observe a clear Zeeman splitting of the orbital states in electron transport measurements through the QD, and find no signature of spin relaxation in our experimental time window, leading to a lower bound on $T_{1}$ of 50 $\mu$s. This lower bound is two to three orders of magnitude longer than spin relaxation times observed in bulk \textit{n}-type GaAs \cite{Kikkawa}, GaAs quantum wells \cite{Ohno} and InAs QDs \cite{Cortez}.

\begin{figure}[ht]
\includegraphics[width=3.4in]{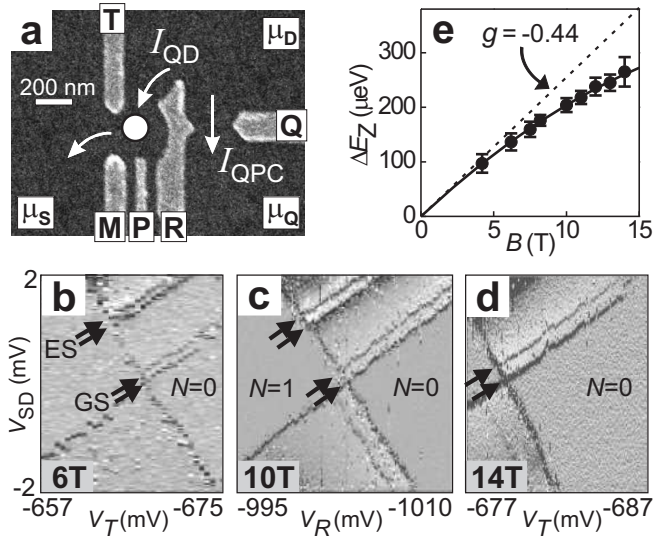}
\caption{\label{Fig 1} (a) Scanning Electron Micrograph of the metallic surface gates \cite{HiddenGates}. Gates $M$, $R$ and $T$ are used to form the quantum dot indicated by a white circle. Additionally, gate $Q$ can be used to form a quantum point contact (QPC). To apply high-frequency signals, gate $P$ is connected to a coaxial cable. Currents through the dot, $I_{QD}$, and through the QPC, $I_{QPC}$, are measured as a function of applied bias voltage, $V_{SD}=(\mu_{S}\!-\!\mu_{D})/e$ and $V_{QD}=(\mu_{Q}\!-\!\mu_{D})/e$ respectively.
(b)-(d) Differential conductance $dI_{QD}/dV_{SD}$ as a function of $V_{SD}$ and gate voltage near the 0$\leftrightarrow$1 electron transition, at parallel magnetic fields of 6, 10 and 14 T. Darker corresponds to larger $dI_{QD}/dV_{SD}$. The zero-field spin degeneracy of both the ground state (GS) and the first orbital excited state (ES) is lifted by the Zeeman energy as indicated by arrows.
(e) Extracted Zeeman splitting $\Delta E_{Z}$ as a function of $B$. At high fields a clear deviation from the bulk GaAs $g$-factor of -0.44 (dashed line) is observed.}
\end{figure}

The quantum dot is defined in a GaAs/Al$_{0.3}$Ga$_{0.7}$As heterostructure, containing a 2DEG 90 nm below the surface with an electron density ${n_{s}=2.9\times 10^{11}}$ cm${^{-2}}$ (Fig. 1a). A magnetic field (0-14 T) is applied parallel to the 2DEG. All measurements are performed in a dilution refrigerator at base temperature \textit{T} = 20 mK.

We tune the device to the few-electron regime and identify the 0$\leftrightarrow$1 electron transition by the absence of further transitions under applied source-drain voltage up to 10 mV. The electron number is confirmed by using the nearby QPC as a charge detector \cite{Field,Jero,Sprinzak}. We find a charging energy of 2.4 meV and an orbital level spacing of 1.1 meV at $B$ = 0 T.

In a parallel magnetic field, the electron states acquire a Zeeman energy shift, which causes the orbital levels to split by $\Delta E_{Z}\!=\!g \mu_B B$ \cite{Weis}. Figs. 1b-d show stability diagrams \cite{LeoFewEl} around the 0$\leftrightarrow$1 electron transition, measured at \textit{B}$\,$=$\,$6 T, 10 T and 14 T. A clear Zeeman splitting of both ground and first orbital excited state is seen directly in this spectroscopy measurement \cite{Potok}. A least-squares fit of the data to a second-order polynomial, which extrapolates with negligible deviation to the origin, gives
\begin{equation}
	\left|g\right| =(0.43\pm 0.04)-(0.0077\pm0.0020)\:B\:(T) \;,
\end{equation}
similar to early measurements on 2DEGs \cite{Dobers}. If we force the fit to be linear in $B$, we get $\left|g\right|\!=\!0.29\pm0.01$, with a zero-field splitting $\Delta E_{Z,B=0}\!\!=\!\!(34 \pm6) \mu$eV.

Factors which can influence the magnetic field dependence of the $g$-factor include: (1) extension of the electron wavefunction into the Al$_{0.3}$Ga$_{0.7}$As region, where $g\!=\!+0.4$ \cite{Snelling,Awschalom}, (2) thermal nuclear polarization, which decreases the effective magnetic field through the hyperfine interaction \cite{BookOptical}, (3) dynamic nuclear polarization due to electron-nuclear flip-flop processes in the dot, which enhances the effective magnetic field \cite{BookOptical}, and (4) the nonparabolicity of the GaAs conduction band \cite{Snelling}. More experiments are needed to separate these effects, which is outside the scope of this Letter.

The two spin states \ket{\ua} (lowest energy) and \ket{\da} can be used as the basis states of a quantum bit. In order to perform quantum operations and to allow sufficient time for read-out of the quantum bit, it is necessary that the spin excited state \ket{\da} be stable. We investigate this by measuring the relaxation time from \ket{\da} to \ket{\ua}. By applying short pulses to gate $P$, we can modulate the potential of the dot and thus the position of the energy levels relative to the electrochemical potentials of the leads, $\mu_{S}$ and $\mu_{D}$. This enables us to populate the spin excited state \ket{\da} and monitor relaxation to \ket{\ua}.
The applicability of various pulse methods for measuring the spin relaxation time depends on two timescales. If the relaxation rate $W$ ($=\!1/T_{1})$ is at least of the same order as the outgoing tunnel rate $\Gamma_{D}$, i.e. $W\!\geq \!\Gamma_{D}$, we can determine $T_{1}$ by applying single-step pulses. This method has previously been used to measure the relaxation time between orbital levels in a QD ($\sim$10 ns) \cite{FujisawaNature}. In the other limit, $W\!<\!\Gamma_{D}$, a more elaborate method using double-step pulses is needed \cite{FujisawaNature}. We proceed as follows. First, we apply single-step pulses to show that $W\!<\!\Gamma_{D}$. Then we apply double-step pulses to measure $T_{1}$. All data shown are taken at $B$ = 7.5 T, and reproduced at 14 T. At fields below 6 T the Zeeman splitting is too small to be resolved in pulse experiments. The bias voltage is always much smaller than the charging energy, thus allowing at most one electron on the dot.

\begin{figure}[t]
\includegraphics[width=3.4in]{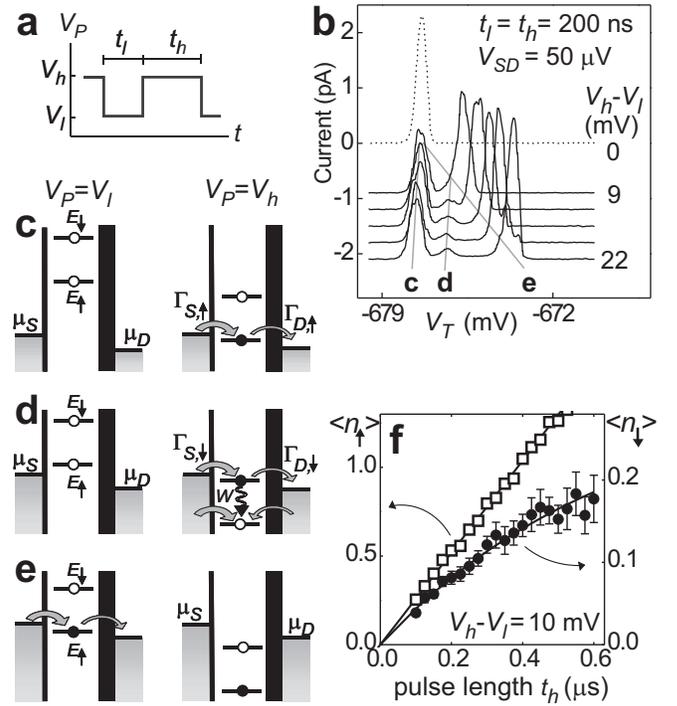}
\caption{One-electron spin relaxation studied using single-step pulses at 7.5T. (a) Schematic waveform of the pulse train (rise/fall time of 0.2 ns). (b) Current traces under applied pulses, offset for clarity.
(c)-(e) Diagrams showing the position of the energy levels during the two phases of the pulse for threee different gate voltage settings, corresponding to the three peaks in (b). (f) Average number of electrons tunneling per cycle (=$I(t_{l}+t_{h})/e$) through the ground state \aver{n_{\ua}}, as in (c), and through the excited state \aver{n_{\da}}, as in (d), vs.  pulse length $t_{h}$. The \aver{n_{\ua}} shows no decay, as expected for a stable current, whereas \aver{n_{\da}} saturates. However, this saturation is not due to spin relaxation (see text).}
\label{Fig 2}
\end{figure}

The single-step pulses are schematically depicted in Fig. 2a. Fig. 2b shows current traces for different amplitudes of the pulses. Transport of electrons through the ground state takes place when \ket{\ua} lies in the bias window (i.e. $\mu_S\!>\!E_{\ua}\!>\!\mu_D$). When we apply single-step pulses, this condition is met at two different values of the gate voltage $V_T$ and therefore the Coulomb peak splits in two. Fig. 2c shows the positions of the energy levels during the two phases of the pulse for the left peak in Fig. 2b. Here, electrons flow from source to drain during the ``high'' phase of the pulse. Similarly, Fig. 2e corresponds to the right peak in Fig. 2b, where ground state transport occurs during the ``low'' phase of the pulse. When the pulse amplitude exceeds the Zeeman splitting ($\approx$160 $\mu$eV), an extra current peak becomes clearly visible. This peak is due to transient transport via the spin-down state \ket{\da} during the ``high'' phase of the pulse (Fig. 2d).
The transient current flows until \ket{\ua} becomes occupied and Coulomb blockade prohibits other electrons to enter the dot. Occupation of \ket{\ua} can happen either via tunneling of an electron from the leads into \ket{\ua} when the dot is empty, or by spin relaxation from \ket{\da} to \ket{\ua}. For both these processes, the probability to have occurred increases with time. Therefore, the number of electrons tunneling via \ket{\da} per cycle, \aver{n_{\da}}, saturates with increasing pulse length $t_{h}$. In particular, if the tunnel rate $\Gamma_{S}$ through the incoming barrier is much larger than the tunnel rate $\Gamma_{D}$ through the outgoing barrier, i.e. $\Gamma_{S}\!\gg\!\Gamma_{D}$ \cite{Asymmetry}, it can be shown that \cite{FujisawaPRB}

\begin{equation}
	<\!n_{\da}\!\!>\: \simeq \:A \Gamma_{D,\da} (1-e^{-Dt_{h}})/D \; ,
	\label{naver}
\end{equation}
where $A \simeq \Gamma_{S,\da}/(\Gamma_{S,\ua}+\Gamma_{S,\da})$ is the injection efficiency into \ket{\da}, and $\Gamma_{D,\da}$ is the tunnel rate from \ket{\da} to the drain (see Fig. 2c-d). The saturation rate $D$ is the sum of $W$, the spin relaxation rate from \ket{\da} to \ket{\ua}, and $(1\!\!-\!\!A)\Gamma_{D,\da}$, which accounts for direct tunneling into \ket{\ua}:
\begin{equation}
	D=W+(1\!\!-\!\!A)\Gamma_{D,\da} \;.
	\label{Dtotal}
\end{equation}
By measuring \aver{n_{\da}} for different pulse widths $t_{h}$, we can find $D$ and $A \Gamma_{D,\da}$ using Eq. (\ref{naver}). Together with the value of $A$, which can be extracted from large-bias measurements without pulses, we can determine the spin-relaxation rate $W$=1/$T_{1}$ via Eq. (\ref{Dtotal}).

In Fig. 2f we show the average number of tunneling electrons per cycle for the stable current, \aver{n_{\ua}}, and for the transient current, \aver{n_{\da}}. Clearly, \aver{n_{\ua}} increases linearly with pulse length, whereas \aver{n_{\da}} saturates, as expected. From fitting \aver{n_{\da}} to Eq. (\ref{naver}) we find $D$=(1.5 $\pm$ 0.2) MHz and $A \Gamma_{D,\da}$=(0.47 $\pm$ 0.09) MHz. Furthermore, $A$=(0.28 $\pm$ 0.05), leading to $(1\!\!-\!\!A)\Gamma_{D,\da}$=(1.2 $\pm$ 0.3) MHz and $W$=(0.30$\pm$ 0.35) MHz. Averaging over similar measurements, using different tunnel rates and $t_{l}$, leads to $W$=(0.20$\pm$ 0.25) MHz. 

We conclude that the spin relaxation rate ($W\!\!<\!$ 0.5 MHz) is much smaller than the tunnel rates ($\Gamma_{S}\!\gg\! \Gamma_{D}\!\approx\!$ 1.6 MHz). This means that the decay of the transient current is dominated by direct injection into \ket{\ua}, and therefore the single-step pulse method can only provide a weak lower bound on $T_{1}$. To circumvent this, we decouple the read-out stage from the relaxation stage by inserting an extra pulse step. This way, an electron can only tunnel out of the dot \textit{after} the waiting time, enabling us to directly measure the relaxation probabilities as a function of waiting time \cite{FujisawaNature}, as explained below.

\begin{figure}[t]
\includegraphics[width=3.4in, clip=true]{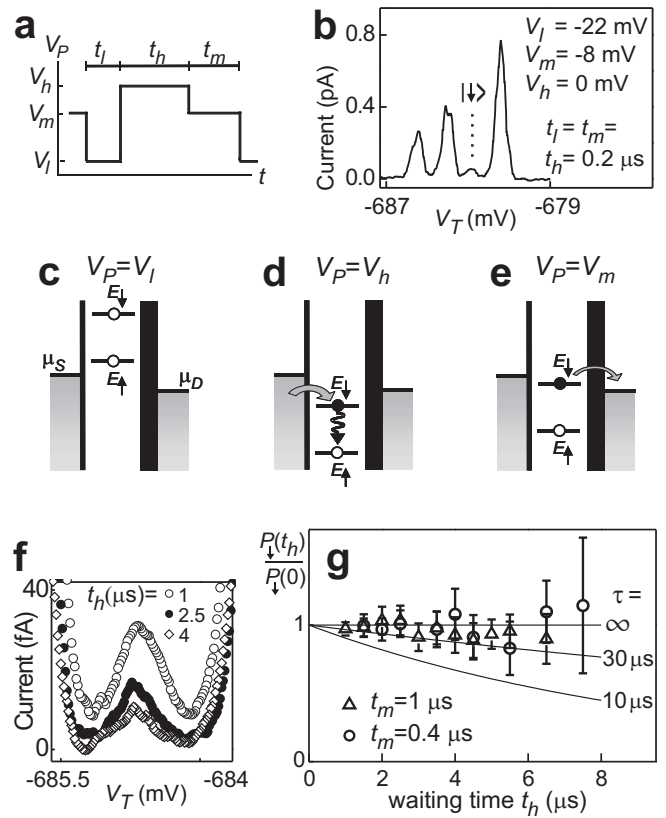}
\caption{\label{Fig 3} One-electron spin relaxation studied using double-step pulses at 7.5T.
(a) Schematic waveform of the pulse train (rise/fall time of 1.5 ns).
(b) Typical pulse-excited current trace. The three main peaks correspond to a stable current flowing via \ket{\ua} when \ket{\ua} is in the bias window during one of the three stages of the waveform. The small peak is due to transient current via \ket{\da} for $V_{P}=V_{m}$ \cite{HiddenPeak}.
(c)-(e) Diagrams depicting the energy levels during the three stages of the pulse for the \ket{\da}-peak shown in (b). (c) The dot is emptied during a time $t_{l}$. (d) Both \ket{\ua} and \ket{\da} lie below the electrochemical potentials of the leads and an electron can tunnel into the \ket{\da}; other possible tunnel processes are not indicated since they do not contribute to the current (see text). We allow the electron to relax for a time $t_{h}$. (e) Now \ket{\da} lies in the bias window. Only if the electron has spin-down it can tunnel out and contribute to current.
(f) Averaged \ket{\da} current peaks for $t_{h}$=1, 2.5 and 4 $\mu$s with $t_{m}$=0.4 $\mu$s (for data in (f) and (g) $t_{l}$=$t_{h}$). (g) The probability $P_{\da}(t_{h})/P_{\da}(0)$ that the spin did \textit{not} decay during the waiting time $t_{h}$.}
\end{figure}

The schematic waveform of the double-step pulses is shown in Fig. 3a. Applying these pulses results in current traces as in Fig. 3b. Figs. 3c-e depict the energy levels for the \ket{\da} current peak indicated in Fig. 3b at the three different stages of the pulse cycle. First the dot is emptied (Fig. 3c). In the second stage (Fig. 3d), an electron tunnels into either \ket{\da} or \ket{\ua}. Again, due to the charging energy only one electron can occupy the dot. The probability that it enters \ket{\da}, $A$, does not depend on the pulse lengths, which are the only parameters we change. If the electron entered \ket{\da}, the probability that it has \textit{not} relaxed to \ket{\ua} after $t_{h}$ is exp(-$t_{h}$/$T_{1}$) (we assume exponential decay). 
Finally (Fig. 3e), if the electron is in \ket{\da}, it can tunnel out, but only to the drain. In contrast, if the electron is in \ket{\ua}, it can tunnel out to either the source or the drain when the cycle is restarted (Fig. 3c). Similarly, electrons entering the dot originate from the source or the drain (Fig. 3d). Assuming that $\Gamma_S/\Gamma_D$ is constant throughout the cycle, the average current generated by electrons leaving the dot during the "low" phase of the pulse train (Fig. 3c) is zero. Therefore the current only consists of electrons that entered \ket{\da} and have not relaxed during $t_h$:

\begin{equation}
	I = e f_{rep} <\!n_{\da}\!\!> \:=\: e f_{rep} C A \:e^{(-t_h/T_{1})},
\end{equation}
where $f_{rep}$ is the pulse repetition frequency and $C$ a constant accounting for the tunnel probability in the read-out stage. We determine \aver{n_{\da}} for different $t_{h}$. Normalized to the value for $t_{h}$=0, it is a direct measure of spin relaxation: 
\begin{equation}
	\frac{<\!n_{\da}\!\!>_{t_{h}=t}}{<\!n_{\da}\!\!>_{t_{h}=0}}=\frac{C A\: e^{(-t/T_{1})}}{C A \:e^{(-0/T_{1})}} = \frac{P_{\da}(t)}{P_{\da}(0)} = e^{(-t/T_{1})} \; .
	\label{decay}
\end{equation}
To be able to extract reliable peak heights from the very small currents, we average over many traces. Examples of averaged curves are shown in Fig. 3f for $t_{h}$=1, 2.5 and 4 $\mu$s. In Fig. 3g, data extracted from these and similar curves are plotted as a function of $t_{h}$, up to 7.5 $\mu$s. Longer waiting times result in unmeasurably small currents ($I\propto 1/t_{h}$). The two data sets shown were taken with different gate settings (and thus different tunnel rates) and different $t_{m}$. As a guide to the eye, lines corresponding to an exponential decay with decay times $\tau\!$ = $\!10\, \mu$s, $\tau\!$ = $\!30\, \mu$s and $\tau\!$ = $\!\infty$ are included. There is no clear decay visible. We fit the data in Fig. 3g and similar data, and average the resulting relaxation rates. From an error analysis we find a lower bound of $T_{1}>$ 50 $\mu$s. We emphasize that, since we do not observe a clear signature of relaxation in our experimental time window, $T_{1}$ might actually be much longer.

The lower bound we find for $T_{1}$ is much longer than the time needed for read-out of the quantum bit using proposed spin-to-charge conversion schemes \cite{Lieven}. In these schemes, spin-dependent tunneling events correlate the charge on the dot to the initial spin state. 
A subsequent charge measurement thus reveals information on the spin. This can de done in our device using the QPC located next to the QD (see Fig. 1a) \cite{Jero}.

An interesting question is how much the stability of the spin states is affected by such charge measurements. We have studied this by sending a large current through the QPC, set at maximum charge sensitivity, and repeating the $T_{1}$ measurements. The drain lead is shared by the QPC- and the QD-current, which causes some peak broadening and limits the experimental window. However, even for a very large current of $\sim$20 nA through the QPC ($\mu_{Q}\!\!\,-\!\!\,\mu_{D}\!=\!500\,\mu eV$), we still do not find a measurable decay of the spin. For comparison, we can measure the charge on the QD within 50 $\mu$s using a QPC current of only 10 nA \cite{JeroUnpublished}. 
Taking these measurements together shows that, by using spin-to-charge conversion, it should be possible to perform single-shot spin readout in this device.

We thank T. Fujisawa, S. Tarucha, T. Hayashi, T. Saku, Y. Hirayama, S.I. Erlingsson, Y.V. Nazarov, O.N. Jouravlev, S. De Franceschi, D. Gammon and R.N. Schouten for discussions and help. This work was supported by the DARPA-QUIST program.


\begin{references}
\bibitem{HiddenGates}
For clarity, we have hidden the surface gates that are not used in the experiments. The complete gate structure can be found in Ref. \cite{Jero}.
\bibitem{Asymmetry}
The condition $\Gamma_{S}\gg\Gamma_{D}$ is easily achieved by tuning the different gate voltages.
\end{references}
\end{document}